\documentstyle[eqsecnum,aps]{revtex}
\begin{document}
\hspace{-15mm}
\vspace{-10.0mm} 

\thispagestyle{empty}
{\baselineskip-4pt
\font\yitp=cmmib10 scaled\magstep2
\font\elevenmib=cmmib10 scaled\magstep1  \skewchar\elevenmib='177
\leftline{\baselineskip20pt
\hspace{12mm} 
\vbox to0pt
   { {\yitp\hbox{Osaka \hspace{1.5mm} University} }
     {\large\sl\hbox{{Theoretical Astrophysics}} }\vss}}

\rightline{\large\baselineskip20pt\rm\vbox to20pt{
\baselineskip14pt
\hbox{OU-TAP 46}
\hbox{KUNS 1419}
\vspace{2mm}
\hbox{May 1997}\vss}}

\vspace{2cm} 

\begin{center}
{\Large\bf Gravitational Radiation Reaction \\ 
to a Particle Motion II} \\ 
\medskip
{\Large\it - Spinning Particle -}
\end{center}
\bigskip

\centerline{\large Yasushi Mino,$^{1,2}$\footnote{Electronic 
address: mino@vega.ess.sci.osaka-u.ac.jp}
 Misao Sasaki$^1$\footnote{Electronic 
address: misao@vega.ess.sci.osaka-u.ac.jp}
and Takahiro Tanaka$^1$\footnote{Electronic 
address: tama@vega.ess.sci.osaka-u.ac.jp}}
\bigskip
\begin{center}{\em $^1$Department of Earth and Space Science, 
Graduate School of Science} \\
{\em  Osaka University, Toyonaka 560, Japan}\\
{\em $^2$Department of Physics, Faculty of Science, 
Kyoto University, Kyoto 606-01, Japan}
\end{center}

\bigskip

\begin{abstract}
We discuss the leading order correction 
to the equation of motion of a particle with spin on an arbitrary
spacetime. 
A particle traveling in a curved spacetime 
is known to trace a geodesic of the background spacetime 
if the mass of the particle is negligibly small. 
For a spinning particle, it is known that there appears
a term due to the coupling of the spin and the Riemann tensor of
the background spacetime.
Recently we have found the equation of motion of a non-spinning
particle which includes the effect of gravitational radiation reaction.
This paper is devoted to discussion of a consistent derivation 
of the equation of motion which is corrected 
both by the spin-Riemann coupling and 
the gravitational radiation reaction. 

\noindent
PACS number(s): 04.30.Db, 04.25.Nx
\end{abstract}


\section{Introduction} 
The gravitational waves from an inspiralling compact binary is 
one of the most promising sources expected to be detected 
by the near-future interferometric gravitational wave detectors 
such as LIGO, VIRGO, GEO, TAMA and LISA\cite{Thorne,LISA}.  
In order to extract the information of a binary
from the last inspiralling stage, 
we need accurate theoretical templates of 
the gravitational wave forms\cite{Last}. 
As an approach, perturbations of a black hole by an orbiting 
point particle have been studied\cite{bhp}. 

Most of the previous papers considered the case that 
the orbiting point particle is structureless and 
is characterized only by a mass parameter. 
In those papers, the point particle is defined 
by using the Dirac delta function, 
therefore the induced perturbation diverges at the particle. 
This leads to some conceptual problems in the perturbation study; 
1) how far we can incorporate the strong non-linearity 
around the particle, 
and 2) how definitely we can define the motion of the particle 
though the perturbed metric diverges at the particle. 

In a previous paper\cite{Mino1} we have considered these problems
in the case of a non-spinning particle, 
extensively using the technique of the matched asymptotic expansion 
that had been developed by many authors 
(e.g., D'Eath \cite{Death} and Thorne and Hartle \cite{Thorne1}). 
The matched asymptotic expansion is a technique to match 
physical quantities in different perturbation schemes. 
In the case of a non-spinning particle, the procedure goes
as follows. We first construct the metric in two different regions.
We assume the metric around a particle (body's neighborhood\cite{Thorne1}) 
to be approximated by a Schwarzshild geometry plus its
perturbation describing tidal distortions due to the background curvature 
(internal scheme\cite{Mino1} or body expansion\cite{Thorne1}). 
In this region the metric is expanded in powers of $r/L$,
where $r$ is the circumference radius measured 
in the local inertial frame of the background and 
$L$ is the characteristic curvature scale of the background spacetime. 
The metric far from the point particle (external 
universe\cite{Thorne1}) is approximated by a given background
spacetime plus the perturbation generated by a point source 
(external scheme\cite{Mino1} or external-universe 
expansion\cite{Thorne1}). 
The metric in this region is expanded in powers of $Gm/r$, 
where $m$ is the mass of the particle. 
The particle is presumed to be compact and isolated in the sense 
that $Gm$ is far smaller than $L$.
Then the metrics constructed in different expansion schemes are
matched in the overlapping region (buffer region\cite{Thorne1}) 
where both expansion schemes are valid.
In practice, the metrics are expanded in powers of both $r/L$ and $Gm/r$
and they are matched by imposing appropriate coordinate conditions.
As we take a Schwarzshild geometry  
as the lowest order approximation of the solution around the particle,  
non-linearity can be fully taken into account there. 

In \cite{Mino1}, the equation of motion has been derived 
from the condition that the above asymptotic matching 
of the metric can be consistently done. 
In the internal scheme, 
the background spacetime is a Schwarzschild 
geometry which has spherical symmetry. This enabled us to
decompose the metric perturbation in the internal scheme 
in terms of the tensor spherical harmonics. 
The important step was to require 
that the translational gauge modes should vanish 
in the internal scheme, 
which corresponds to fixing the center of mass of the particle. 
Equating each component of the metrics derived 
in the internal and external schemes in the overlapping region, 
we have obtained the equation of motion of the center of mass 
in the backgound spacetime in the external scheme
which includes the effect of gravitational radiation reaction. 
It should be noted that Quinn and Wald have recently obtained 
the same equation of motion for a non-spinning particle
by an axiomatic approach without resorting to the matched 
asymptotic expansion\cite{axiom}. However, it is beyond the scope 
of this paper to discuss the physical meaning of their axioms in 
the framework of the matched asymptotic expansion.

Having shown that the asymptotic matching can be consistently done 
with the external metric generated by a non-spinning point particle, 
we have clarified that the expression of the energy momentum tensor 
for a point particle used in computing 
the gravitational radiation reaction to the orbit in previous
works\cite{bhp} can be regarded as a Schwarzschild black hole
or any compact object provided that it is kept 
sufficiently spherically symmetric.
Since it is unnecessary to evaluate the external metric 
at the location of the particle, 
we do not encounter the divergence even if we use the 
delta function source.  
Applying the asymptotic matching, we obtain the equation of motion 
of the origin of the internal frame  
with respect to the background spacetime. 
And hence the equation of motion is understood in a well-defined manner.
As a result, we found that  
the strong equivalence principle holds to $O\left((Gm/L)^2\right)$
for a non-spinning compact body in the limit $Gm/L\ll1$.
Here compact body means that it has no typical length scale  
other than its Schwarzshild radius. 

On the other hand, the post-Newtonian calculation shows that 
the spin of the particle modifies the gravitational wave forms 
of a realistic binary system\cite{Kidder}, 
which motivates us to study the problem of incorporating
a spinning particle into the perturbation theory. 
For this purpose, we need to know the correction 
to the equation of motion due to spin and the 
energy momentum tensor of a spinning particle. 
 
The effect of the spin to the equation of motion was discussed 
by many authors in various ways\cite{Papa,Dixon}.
Dixon\cite{Dixon} found the covariant description 
of the energy momentum tensor of an extended body 
and the equation of motion of it. 
The resultant equation of motion shows
that the coupling of the spin and the Riemann curvature 
affects the orbit of the body. 
However, the Dixon's equation of motion is written in terms of
the full metric which can be obtained only after the Einstein equations
are completely solved. 
On the other hand, what we are interested in is to describe
the motion of a compact body and the effect of radiation reaction 
in a given external background. 
For this purpose, further reduction is necessary. 
Furthermore, Dixson's equation of motion does not seem to be
applicable to a strongly self-gravitating body like a
black hole in a strict sense. 

The effect of the spin was also discussed, for example, 
in D'Eath\cite{Death} and Thorne and Hartle\cite{Thorne1}, 
in which the strong self-gravity of the body is taken into account
by using the technique of matched asymptotic expansion. 
Especially Thorne and Hartle\cite{Thorne1} established 
the framework of matched asymptotic expansion of the metric 
around the particle. 
Using the metric thus derived, 
they integrated non-covariant conservation laws 
written in the inertial frame of the background spacetime
(body's local asymptotic rest frame in \cite{Thorne1}) 
on the world tube enclosing the particle. 
They found that these integrations contain the
 `laws of motion and precession' 
which can be converted into the equations of motion and spin
when an approximate solution of the Einstein equations 
of the surrounding spacetime is given. 
The laws of motion indicate that the geodesic motion is 
corrected by the coupling of the spin of the body 
and the Riemann curvature of the surrounding spacetime. 
As a specific example of a system, they considered the case 
of two Kerr black holes orbiting each other. 
They assumed that 
the separation of these two black holes is large enough 
and derived the equations of motion and spin
which are accurate up through the 1.5th post-Newtonian order. 
Since the radiation reaction force is known to 
give rise from 2.5 post-Newtonian order, 
the correction due to the radiation reaction derived in \cite{Mino1} 
did not appear in \cite{Thorne1}.

The purpose of this paper is 
to derive the correction to the equation of motion 
due to the radiation reaction 
as well as to the spin-Riemann coupling in a unified manner. 
Again we use the technique of matched asymptotic expansion 
in order to construct the metric. 
Recently we have discussed 
the perturbation due to a spinning point particle\cite{Mino2} 
and calculated the gravitational waves from a spinning particle
orbiting a Kerr black hole up through the 2.5th post-Newtonian order.
In this calculation, we have assumed that the spin contribution 
to the energy momentum tensor is given by a derivative of the 
delta function. 
The result agrees with the one obtained 
by the standard post-Newtonian calculation\cite{Kidder} in a suitable
limit.
Hence, we assume that metric far from a spinning particle is also 
approximated by a linear perturbation of a general background spacetime 
generated by the delta function source, 
and we assume that the metric around the particle is 
approximated by that of the Kerr geometry with perturbations,
which should/will be justified by the consistency condition of the matching.

The next step is to extract the information 
of the motion of the particle from the metric thus constructed. 
At this point 
we have two methods to obtain the equation of motion; 
the consistency condition of the matching 
with an appropriate choice of gauge in the internal scheme\cite{Mino1}, 
or the use of non-covariant conservation laws 
in the inertial frame of the background spacetime\cite{Thorne1}. 
The former method developed in \cite{Mino1} 
seems difficult to apply to the present case. 
This is because we do not know a method to distinguish 
the translational gauge modes of the metric perturbation 
in a Kerr background. Hence we cannot fix the center of mass condition 
in the internal scheme when matching the metrics in the internal and 
external schemes in the overlapping region. 
Thus we adopt the latter method and integrate the conservation laws 
as in \cite{Thorne1} to derive the equation of motion.
Relating to the difficulty of the former method, 
we cannot verify the consistency of 
matching to the same extent as we could in the case 
of non-spinning particle. Thus we simply assume it. 

This paper is organized as follows. 
In section 2, 
we construct the metric in the overlapping region (buffer region), 
using the covariant expansion method of 
the tensor Green function\cite{Mino1}. 
We suppose the reader is familiar with the concept of 
`bi-tensors'\cite{DeWitt} and skip the computation of the 
tensor Green function\cite{Mino1}. 
In section 3, we derive the equation of motion of the spinning 
point particle. 
Section 4 summarizes the result.


\section{Matched Asymptotic Expansion}

Following the procedure given in \cite{Mino1}, 
we first discuss the external scheme of the metric. 
We have a general background metric $g_{\mu\nu}(x)$ 
which satisfies the vacuum Einstein equations 
around the particle in the external scheme. 
We compute its linear perturbation with the Green function 
method. 
The perturbation in the external scheme is constructed 
with an assumption on the perturbation source. 
(See Eq. (\ref{eq:source}) below.) 
We will discuss the consistency of this assumption later
when we compute the metric in the overlapping region.

We put the metric $g_{\mu\nu}(x)+h_{\mu\nu}(x)$ into the 
Einstein equations and solve the linearized 
equations for $h_{\mu\nu}$ assuming that the background 
metric $g_{\mu\nu}$ satisfies the vacuum Einstein equations 
around the particle of interest\footnote{In this derivation, 
we follow the notation used in \cite{DeWitt}. 
We assign the indices $\alpha,\beta,\gamma,\delta$ 
for the point on the particle trajectory $z(T)$, and 
the indices $\mu,\nu,\xi,\rho$ for the field point $x$}. 
We introduce the trace-reversed metric perturbation 
$\psi_{\mu\nu}(x)$, and put the covariant harmonic gauge 
condition on it. 
\begin{eqnarray}
\psi_{\mu\nu}(x)
&=& h_{\mu\nu}(x)-{1\over 2} g_{\mu\nu}(x)g^{\xi\pi}(x)h_{\xi\pi}(x) \\
\psi^{\mu\nu}{}_{;\nu} &=& 0 
\end{eqnarray}
We can make use of the tensor Green function 
of the linearized Einstein equations with the retarded boundary 
condition derived in Sec. 2 of \cite{Mino1}. 
Here we assume the metric perturbation is generated 
by the following source term\cite{Mino2}, 
\begin{eqnarray}
T^{\mu\nu}(x)=m\int dT \left\{v^\mu(x,T) v^\nu(x,T) 
{\delta^{(4)}(x-z(T))\over \sqrt{-g}}
-\nabla_\xi\left(S^{\xi(\mu}(x,T)v^{\nu)}(x,T)
{\delta^{(4)}(x-z(T))\over \sqrt{-g}}\right)\right\}, 
\label{eq:source}
\end{eqnarray}
where $v^\mu(x,T)=\bar g^\mu{}_\alpha(x,z(T))\dot z^\alpha(T)$ 
and $S^{\mu\nu}(x,T)=\bar g^\mu{}_\alpha(x,z(T)) 
\bar g^\nu{}_\beta(x,z(T))S^{\alpha\beta}(T)$. 
$\bar g^\mu{}_\alpha(x,z(T))$ is 
a bivector of parallel displacement 
defined, for example, in Appendix A of \cite{Mino1}, 
and $S^{\alpha\beta}(T)$ is an anti-symmetric tensor 
which is called the spin tensor of the particle 
and is assumed to satisfy the center of mass condition, 
$S_{\alpha\beta}(T)v^\beta(T)=0$. 
A dot over a function means $T$-derivative of it, 
such as $\dot z^\alpha(T)=d z^\alpha/dT$.
The amplitude of the spin tensor $S_{\alpha\beta}(T)$ is 
usually assumed to be of order $m$ in the case of a Kerr black 
hole, which seems to indicate the 2nd order perturbation 
theory is necessary when we want to discuss the spin effect 
of the particle in a consistent expansion with respect to $m$. 
However, we will find it not so if we 
assume that the construction of the metric by the matched asymptotic 
expansion is consistent. 
terms of order $m^2$, the 2nd order perturbation theory is 
unnecessary. 
The exact form of the metric perturbation becomes 
\begin{eqnarray}
\psi^{\mu\nu}(x)&=&
2Gm\Biggl(\biggl[ 
{1\over\dot\sigma(x,z(T))}u^{\mu\nu}{}_{\alpha\beta}(x,z(T)) 
\dot z^\alpha(T) \dot z^\beta(T) 
\nonumber \\ && \qquad \qquad 
+{\ddot\sigma(x,z(T))\over\dot\sigma^3(x,z(T))} 
u^{\mu\nu}{}_{\alpha\beta}(x,z(T))\sigma_{;\gamma}(x,z(T)) 
S^{\gamma\alpha}(T)\dot z^\beta(T) 
\nonumber \\ && \qquad \qquad 
+{1\over\dot\sigma(x,z(T))}u^{\mu\nu}{}_{\alpha\beta;\gamma}(x,z(T)) 
S^{\gamma\alpha}(T) \dot z^\beta(T) 
\nonumber \\ && \qquad \qquad 
-{1\over\dot\sigma^2(x,z(T))}{d\over dT}\left(
u^{\mu\nu}{}_{\alpha\beta}(x,z(T))\sigma_{;\gamma}(x,z(T))
S^{\gamma\alpha}(T) \dot z^\beta(T) \right)
\nonumber \\ && \qquad \qquad 
+{1\over\dot\sigma(x,z(T))}v^{\mu\nu}{}_{\alpha\beta}(x,z(T))
\sigma_{;\gamma}(x,z(T)) S^{\gamma\alpha}(T)\dot z^\beta(T) 
\biggr]_{T=T_{Ret}(x)} 
\nonumber \\ && \qquad \qquad 
-\int^{T_{Ret}(x)}_{-\infty}
dT \left(v^{\mu\nu}{}_{\alpha\beta}(x,z(T))
\dot z^\alpha(T)\dot z^\beta(T)
+v^{\mu\nu}{}_{\alpha\beta;\gamma}(x,z(T))
S^{\gamma\alpha}(T)\dot z^\beta(T)\right)
\Biggr),
\label{eq:metper} 
\end{eqnarray} 
where $\sigma(x,z)$ is a half the squared geodetic interval 
between $x$ and $z$ and is defined, for example, 
in Appendix A of \cite{Mino1},
$u^{\mu\nu\alpha\beta}(x,z)$ and $v^{\mu\nu\alpha\beta}(x,z)$ 
appear in the expression for the tensor Green function
and are defined in Sec. 2 of \cite{Mino1}, and
$T_{Ret}(x)$ is the retarded time of the particle which is
a scalar function determined by $\sigma\left(x,z(T_{Ret})\right)=0$ 
with the condition that $z(T_{Ret})$ is on the past light 
cone of $x$. 

We next consider the metric in the overlapping region, in which 
the metric is expanded in powers of $Gm/r$ and $r/L$. 
We first introduce coordinates 
$\{ T, X^i\}=\{X^a\}, \quad (i=1,2,3;a=0,1,2,3)$\footnote{
We assign the indices, $a,b,c,d$, for the spacetime coordinates 
i.e. $0,1,2,3$, while the indices, $i,j,k,l$, 
for the spatial coordinates i.e. $1,2,3$. 
Minkowski and Kronecker summention conventions are taken.}, 
such that they become those introduced in \cite{DeWitt} 
when taking the test particle limit $m\rightarrow 0$ :
\begin{eqnarray}
\lim_{m \rightarrow 0}
\left(\sigma_{;\alpha}(x,z(T))+e_{\alpha i}(T) X^i \right)&=& 0,
\label{eq:coord} \\ 
\lim_{m \rightarrow 0}\dot z^2(T) &=& -1, \label{eq:coord1} \\ 
\lim_{m \rightarrow 0}\dot z^\alpha(T)e_{\alpha i}(T) &=& 0,
\label{eq:coord2} \\ 
\lim_{m \rightarrow 0}e^\alpha{}_i(T)e_{\alpha j}(T) &=& \delta_{ij}.
\label{eq:coord3} 
\end{eqnarray}
We assume that $X^i X^i/L^2$ is small enough in the overlapping region
so that the coordinate transformation $x^\alpha =x^\alpha (X^a)$ is
single valued\cite{DeWitt}.
Eqs.~(\ref{eq:coord2}) and (\ref{eq:coord3}) show
that $e_{\alpha i}(T)$ are the spatial triad at $z(T)$ 
on the 3-hypersurface normal to $\dot z(T)$.
The coordinate system for finite $m$ shall be determined later. 

We denote the components of the metric whose order of magnitude are 
$O((Gm)^n/L^m)$ by ${}^{(m)}_{(n)}h_{ab}$. 
\begin{eqnarray}
\stackrel{full}{g_{ab}}=\sum_{m,n=0}^{\infty}{}^{(m)}_{(n)}h_{ab}
\end{eqnarray}
We call these terms the $({}^m_n)$-components of the metric\cite{Mino1}.
Simple dimensional analysis shows 
\begin{eqnarray}
{}^{(m)}_{(n)}h_{ab} \sim |X|^{(m-n)}, \label{eq:dim}
\end{eqnarray}
where $|X|=\sqrt{X^i X^i}$.
In the next section, we use the Landau-Lifshitz pseudotensor 
in order to compute the equation of motion, and 
the metric is expanded by taking the flat Minkowski metric $\eta_{ab}$ 
as a background. For its purpose, we raise 
the indices of the $({}^m_n)$-components of the metric by $\eta_{ab}$, 
and we define the trace-reversed $({}^m_n)$-components of the metric 
with respect to the flat Minkowski. 
\begin{eqnarray}
{}^{(m)}_{(n)}\bar h_{ab}={}^{(m)}_{(n)}h_{ab}
-{1\over 2}\eta_{ab}\eta^{cd}{}^{(m)}_{(n)}h_{cd}
\end{eqnarray}

The $({}^m_0)$-components of the metric are obtained from 
the background metric $g_{\mu\nu}(x)$. The background metric 
in the coordinates $\{ X^a \}$ is partly given 
in \cite{Mino1}. A further calculation results in 
\begin{eqnarray} 
g_{\mu\nu}(x)dx^\mu dx^\nu &=& 
\biggl(\dot z^2(T)+2\dot z^\alpha(T){De_{\alpha i} \over dT}(T)X^i 
+{De^\alpha{}_i \over dT}(T){De_{\alpha j} \over dT}(T)X^i X^j
\nonumber \\ && \quad
+R_{\alpha\beta\gamma\delta}(z(T))\dot z^\alpha(T)X^\beta(T)\dot 
z^\gamma(T) 
X^\delta(T)+O(|X|^3)\biggr)dT^2 
\nonumber \\ && 
+2\biggl(\dot z^\alpha(T) e_{\alpha i}(T) 
+e^\alpha{}_i(T){De_{\alpha j} \over dT}(T)X^j 
\nonumber \\ && \qquad \quad 
+{2\over 3}R_{\alpha\beta\gamma\delta}(z(T))e^\alpha{}_i(T)X^\beta(T) 
\dot z^\gamma(T)X^\delta(T)+O(|X|^3)\biggr)dTdX^i 
\nonumber \\ && 
+\biggl(e^\alpha{}_i(T)e_{\alpha j}(T) 
\nonumber \\ && \qquad 
+{1\over 3}R_{\alpha\beta\gamma\delta}(z(T))e^\alpha{}_i(T)X^\beta(T) 
e^\gamma{}_j(T)X^\delta(T)+O(|X|^3)\biggr)dX^i dX^j,
\label{eq:ext1} 
\end{eqnarray} 
where $X^\alpha(T)=e^\alpha{}_i(T)X^i$.
Because of (\ref{eq:dim}), we find 
\begin{eqnarray}
{}^{(0)}_{(0)}h_{00} &=& \dot z^2(T) + O(Gm/L), \label{eq:00tt} \\
{}^{(0)}_{(0)}h_{0i} &=& \dot z^\alpha(T)e_{\alpha i}(T) + O(Gm/L), 
\label{eq:00ti} \\
{}^{(0)}_{(0)}h_{ij} &=& e^\alpha{}_i(T)e_{\alpha j}(T) + O(Gm/L), 
\label{eq:00ij} \\
{}^{(1)}_{(0)}h_{00} &=& 
2 \dot z^\alpha(T) {D e_{\alpha i}\over dT}(T) X^i + O(Gm|X|/L^2), 
\label{eq:10tt} \\
{}^{(1)}_{(0)}h_{0i} &=& 
e^\alpha{}_i(T){D e_{\alpha j}\over dT}(T) X^j + O(Gm|X|/L^2), 
\label{eq:10ti} \\
{}^{(1)}_{(0)}h_{ij} &=& O(Gm|X|/L^2), 
\label{eq:10ij} \\
{}^{(2)}_{(0)}h_{00} &=& R_{\alpha\beta\gamma\delta}(z(T)) 
\dot z^\alpha(T)X^\beta(T)\dot z^\gamma(T)X^\delta(T) 
+ O(Gm|X|^2/L^3), \label{eq:20tt} \\
{}^{(2)}_{(0)}h_{0i} &=& {2\over 3}R_{\alpha\beta\gamma\delta}(z(T)) 
e^\alpha{}_i(T)X^\beta(T)\dot z^\gamma(T)X^\delta(T) 
+ O(Gm|X|^2/L^3), \label{eq:20ti} \\
{}^{(2)}_{(0)}h_{ij} &=& {1\over 3}R_{\alpha\beta\gamma\delta}(z(T)) 
e^\alpha{}_i(T)X^\beta(T)e^\gamma{}_j(T)X^\delta(T) 
+ O(Gm|X|^2/L^3). \label{eq:20ij} 
\end{eqnarray}

As was argued in \cite{Mino1}, the matching condition requires 
that the $({}^0_0)$-components of the metric 
(\ref{eq:00tt})-(\ref{eq:00ij}) should be equal to those of the flat metric 
\begin{eqnarray}
{}^{(0)}_{(0)}h_{ab} &=& \eta_{ab}, 
\end{eqnarray}
which is exactly realized by the coordinate conditions 
(\ref{eq:coord1})-(\ref{eq:coord3}). 

The $({}^1_0)$-components of the metric should vanish 
in order to put the particle 
in the locally rest frame of the background geometry. 
\begin{eqnarray}
{}^{(1)}_{(0)}h_{ab}=0
\end{eqnarray}
{}From (\ref{eq:10tt})-(\ref{eq:10ij}), we thus obtain 
\begin{eqnarray}
\dot z^\alpha(T) {D e_{\alpha i}\over dT}(T) &=& O(Gm/L^2), 
\label{eq:10nn1} \\
e^\alpha{}_i(T){D e_{\alpha j}\over dT}(T) &=& O(Gm/L^2). 
\label{eq:10nn2} 
\end{eqnarray}

{}From (\ref{eq:coord1}), (\ref{eq:coord2}) and (\ref{eq:10nn1}), we get 
the geodesic motion of the particle in the background geometry. 
\begin{eqnarray}
{D \over dT}\dot z^\alpha(T) = O(Gm/L^2) \label{eq:geodesic0} 
\end{eqnarray}
(\ref{eq:10nn1}) and (\ref{eq:10nn2}) show the geodesic transform 
of the spatial triad $e^\alpha{}_i(T)$ in the background geometry. 
\begin{eqnarray}
{De^\alpha{}_i \over dT}(T) = O(Gm/L^2) \label{eq:parallel0} 
\end{eqnarray}

The $({}^m_1)$-components of the metric and 
the $({}^m_2)$-components of the metric due to the spin of the particle 
come from both the background metric $g_{\mu\nu}$ 
and the metric perturbation (\ref{eq:metper}). 
The covariant expansion of the metric perturbation 
due to the monopole particle is already given in \cite{Mino1}. 
Using (\ref{eq:coord}), we can soon get the expression of it 
up to $O(m X)$. 
We further compute the metric perturbation due to the spin 
up to $O(m S |X|^0)$. 
\begin{eqnarray}
\psi_{\mu\nu}(x)dx^\mu dx^\nu &=&
\Biggl\{2Gm\biggl({2\over|X|}
+{1\over 3|X|}R_{\alpha\beta\gamma\delta}(z(T))\dot z^\alpha(T)
X^\beta(T)\dot z^\gamma(T)X^\delta(T)\biggr)
\nonumber \\ && \quad 
-2\left(V_{\alpha\beta}(T)+V_{\alpha\beta\gamma}(T)X^\gamma(T)\right)
\dot z^\alpha(T)\dot z^\beta(T)
+O(m |X|^2, m |S| |X|^0) \Biggr\}dT^2 
\nonumber \\ && 
+2\Biggl\{-2GmR_{\alpha\beta\gamma\delta}(z(T))\dot z^\alpha(T)
X^\beta(T)\dot z^\gamma(T)e^\delta{}_i(T)
\nonumber \\ && \qquad 
-2\left(V_{\alpha\beta}(T)+V_{\alpha\beta\gamma}(T)X^\gamma(T)\right)
\dot z^\alpha(T)e^\beta{}_i(T)
\nonumber \\ && \qquad 
-2G{m \over |X|^3}S_{\alpha\beta}(T)X^\alpha(T)e^\beta{}_i(T)
+O(m |X|^2, m |S| |X|^0) \Biggr\}dTdX^i
\nonumber \\ && 
+\Biggl\{-4Gm|X|R_{\alpha\beta\gamma\delta}(z(T))\dot z^\alpha(T)
e^\beta{}_i(T)\dot z^\gamma(T)e^\delta{}_j(T)
\nonumber \\ && \quad 
-2\left(V_{\alpha\beta}(T)+V_{\alpha\beta\gamma}(T)X^\gamma(T)\right)
e^\alpha{}_i(T)e^\beta{}_j(T)
+O(m |X|^2, m |S| |X|^0) \Biggr\}dX^i dX^j,
\label{eq:metper1} \\ 
&& V^{\alpha\beta}(T) = Gm\int^T_{-\infty}dT'
\Bigl(v^{\alpha\beta}{}_{\alpha'\beta'}(z(T),z(T'))
\dot z^{\alpha'}(T')\dot z^{\beta'}(T')
\nonumber \\ && \qquad \qquad \qquad 
+v^{\alpha\beta}{}_{\alpha'\beta';\gamma'}(z(T),z(T'))
S^{\gamma'\alpha'}(T')\dot z^{\beta'}(T')\Bigr),
\\ 
&& V^{\alpha\beta\gamma}(T) = Gm\int^T_{-\infty}dT'
\Bigl(v^{\alpha\beta}{}_{\alpha'\beta'}{}_{;\gamma}(z(T),z(T'))
\dot z^{\alpha'}(T')\dot z^{\beta'}(T')
\nonumber \\ && \qquad \qquad \qquad 
+v^{\alpha\beta}{}_{\alpha'\beta';\gamma\gamma'}(z(T),z(T'))
S^{\gamma'\alpha'}(T')\dot z^{\beta'}(T')\Bigr).
\end{eqnarray}

We first discuss the $({}^0_1)$-components and 
the $({}^0_2)$-components of the metric. 
{}From (\ref{eq:metper1}), we obtain 
\begin{eqnarray}
{}^{(0)}_{(1)}\bar h_{00}&=& {4Gm \over |X|},
\\
{}^{(0)}_{(1)}\bar h_{0i}&=& 0,
\\
{}^{(0)}_{(1)}\bar h_{ij}&=& 0,
\\
\left[{}^{(0)}_{(2)}\bar h_{00}\right]_{spin} &=& 0,
\\
\left[{}^{(0)}_{(2)}\bar h_{0i}\right]_{spin} &=& -{2 G m \over |X|^3}
S_{\alpha\beta}(T)X^\alpha(T)e^\beta{}_i(T) +O((Gm)^3/L|X|^2), 
\\
\left[{}^{(0)}_{(2)}\bar h_{ij}\right]_{spin} &=& 0. 
\end{eqnarray}
where $[\quad]_{spin}$ means that 
only the spin contribution is taken into account.
In view of the internal scheme, the
$({}^0_n)$-components of the metric describe the background metric, 
thus the consistency condition of the matched asymptotic expansion 
requires that these components realize 
the asymptotic gravitational field of a Kerr black hole. 
\cite{Mino1} argues the $({}^0_1)$-components of the metric fit 
the asymptotic metric of a Schwarzschild black hole 
with mass parameter $m$ and we have shown in \cite{Mino2} 
that the spin contribution to the $({}^0_2)$-components 
of the metric correctly realizes the asymptotic field
of a Kerr black hole. 
Therefore the metric thus constructed in the external scheme 
matches the metric in the internal scheme and 
the use of (\ref{eq:source}) is justified 
in computing the leading order contribution of 
the mass $m$ and the spin $S$. 
The monopole contribution to the $({}^0_2)$-components of the metric 
cannot be computed in the external scheme 
unless we develop the 2nd order perturbation formalism. 
However, the consistency condition of the matched asymptotic expansion 
requires that these components are 
those of the asymptotic expansion of a Schwarzschild black hole. 
Thus we finally obtain 
\begin{eqnarray}
{}^{(0)}_{(2)}\bar h_{00} &=& {(Gm)^2 \over|X|^2}, 
\\
{}^{(0)}_{(2)}\bar h_{0i} &=& -{2 G m \over |X|^3} 
S_{\alpha\beta}(T)X^\alpha(T)e^\beta{}_i(T) +O((Gm)^3/L|X|^2), 
\\
{}^{(0)}_{(2)}\bar h_{ij} &=& {(Gm)^2 \over |X|^2} \left(
-2 \delta_{ij} +{X^i X^j\over |X|^2}\right). 
\end{eqnarray}

We next discuss the $({}^1_1)$-components and 
the $({}^1_2)$-components of the metric. 
As is argued in \cite{Thorne1}, we can take the coordinate system 
such that all the $({}^1_n)$-components vanish as long as 
the $({}^1_0)$-components of the metric vanish. 
In view of the internal scheme, the $({}^1_n)$-components of the metric 
describe a linear perturbation of the background metric. 
This perturbation is induced by the curvature of the external universe, 
i.e. by the $({}^1_0)$-components. However, we have seen that the
$({}^1_0)$-components should vanish because of the coordinate condition.
Thus the $({}^1_n)$-components of the metric have no physical mode, 
and we can set them zero with a suitable choice of coordinates. 
As discussed by Thorne and Hartle\cite{Thorne1}, it is essential to put
them zero for reducing the uncertainties in the definitions of the mass,
momentum and spin of the particle when deriving the equation of motion.
Hence we require ${}^{(1)}_{(n)}h_{ab}=0$.
{}From (\ref{eq:ext1}) and (\ref{eq:metper1}), we obtain 
$({}^1_1)$-components of the metric. 
\begin{eqnarray}
{}^{(1)}_{(1)}\bar h_{00}&=& 
{1\over 2}\left(e^\alpha{}_i(T) e_{\alpha i}(T) +\dot z^2(T) -2\right) 
\nonumber \\ && 
-2 V_{\alpha\beta}(T) \dot z^\alpha(T) \dot z^\beta(T) 
+O((Gm)^2/L^2)
\label{eq:11tt}
\\
{}^{(1)}_{(1)}\bar h_{0i}&=& \dot z^\alpha(T) e_{\alpha i}(T) 
\nonumber \\ && 
-2 V_{\alpha\beta}(T) \dot z^\alpha(T) e^\beta{}_i(T) 
+O((Gm)^2/L^2)
\label{eq:11ti}
\\
{}^{(1)}_{(1)}\bar h_{ij}&=& 
\left(e^\alpha{}_i(T) e_{\alpha j}(T) -\delta_{ij}\right)-{1\over 2}
\delta_{ij}\left(e^\alpha{}_k(T) e_{\alpha k}(T) -\dot z^2(T) -4\right)
\nonumber \\ && 
-2 V_{\alpha\beta}(T) e^\alpha{}_i(T) e^\beta{}_j(T) 
+O((Gm)^2/L^2)
\label{eq:11ij}
\end{eqnarray}
We then obtain the coordinate conditions which satisfy 
${}^{(1)}_{(1)}\bar h_{ab} = 0$. 
\begin{eqnarray}
\sigma_{;\alpha}(x,z(T)) + e_{\alpha i}(T) X^i &=& 0 \\ 
\dot z^2(T) &=& -1 +V_{\alpha\beta}(T) 
\left(g^{\alpha\beta}(z(T)) +2 \dot z^\alpha(T)\dot z^\beta(T)\right) 
+O((Gm)^2/L^2) 
\label{eq:Ltt} \\ 
\dot z^\alpha(T) e_{\alpha i}(T) &=& 
-2 V_{\alpha\beta}(T) \dot z^\alpha(T) e^\beta{}_i(T) +O((Gm)^2/L^2) 
\label{eq:Lti} \\ 
e^\alpha{}_i(T)e_{\alpha j}(T) &=& \delta_{ij} +V_{\alpha\beta}(T)
\left(-\delta_{ij} g^{\alpha\beta}(z(T))
+2e^\alpha{}_i(T)e^\beta{}_j(T)\right) +O((Gm)^2/L^2) 
\label{eq:Lij} 
\end{eqnarray}
Setting the $({}^1_2)$-components of the metric zero will give us
the coordinate conditions 
to the order $O((Gm)^2/L^2)$, but we do not need them for the present
purpose. 

The metric derived by the matched asymptotic expansion 
is summarized as follows. 
\begin{eqnarray}
{}^{(0)}_{(0)}h_{ab}&=&\eta_{ab}\,,
\\
{}^{(1)}_{(0)}\bar h_{ab}&=&0 \,,
\\
{}^{(2)}_{(0)}\bar h_{00}&=& {2\over 3}R_{\alpha\beta\gamma\delta}(z(T)) 
\dot z^\alpha(T) X^\beta(T) \dot z^\gamma(T) X^\delta(T) 
\nonumber \\ && 
+O(Gm|X|^2/L^3) \,,
\\
{}^{(2)}_{(0)}\bar h_{0i}&=& {2\over 3}R_{\alpha\beta\gamma\delta}(z(T))
e^\alpha{}_i(T) X^\beta(T) \dot z^\gamma(T) X^\delta(T) 
\nonumber \\ && 
+O(Gm|X|^2/L^3) \,,
\\
{}^{(2)}_{(0)}\bar h_{ij}&=& {1\over 3}R_{\alpha\beta\gamma\delta}(z(T))
X^\beta(T) X^\delta(T) \left(e^\alpha{}_i(T) e^\gamma{}_j(T) 
+\delta_{ij}\dot z^\alpha(T) \dot z^\gamma(T)\right) 
\nonumber \\ && 
+O(Gm|X|^2/L^3) \,,
\\
{}^{(0)}_{(1)}\bar h_{00}&=& {4Gm \over |X|} \,,
\\
{}^{(0)}_{(1)}\bar h_{0i}&=& 0 \,,
\\
{}^{(0)}_{(1)}\bar h_{ij}&=& 0 \,,
\\
{}^{(1)}_{(1)}\bar h_{ab}&=& 0 \,,
\\
{}^{(2)}_{(1)}\bar h_{00}&=& 
\dot z^\alpha(T) {D\over dT}e_{\alpha i}(T) X^i(T)
\nonumber \\ && 
+{10 G m \over 3 |X|}R_{\alpha\beta\gamma\delta}(z(T)) 
\dot z^\alpha(T) X^\beta(T) \dot z^\gamma(T) X^\delta(T) 
\nonumber \\ && 
-2 V_{\alpha\beta\gamma}(T) \dot z^\alpha(T) \dot z^\beta(T) 
X^\gamma(T) 
+O((Gm)^2 |X|/L^3) \,,
\\
{}^{(2)}_{(1)}\bar h_{0i}&=& 
e^\alpha{}_i(T) {D\over dT}e_{\alpha j}(T) X^j
\nonumber \\ && 
+2 G m R_{\alpha\beta\gamma\delta}(z(T)) X^\beta(T) \dot z^\gamma(T) 
\left(-\dot z^\alpha(T) e^\delta{}_i(T) 
+{2 \over 3 |X|}e^\alpha{}_i(T) X^\delta(T)\right)
\nonumber \\ && 
-2 V_{\alpha\beta\gamma}(T)\dot z^\alpha(T) e^\beta{}_i(T) X^\gamma(T) 
+O((Gm)^2 |X|/L^3) \,,
\\
{}^{(2)}_{(1)}\bar h_{ij}&=& 
\delta_{ij}\dot z^\alpha(T) {D\over dT}e_{\alpha k}(T) X^k
\nonumber \\ && 
+2 G m R_{\alpha\beta\gamma\delta}(z(T)) 
\Bigl({1\over 3|X|}X^\beta(T) X^\delta(T) 
\left(e^\alpha{}_i(T) e^\gamma{}_j(T) 
-\delta_{ij}\dot z^\alpha(T) \dot z^\gamma(T) \right) 
\nonumber \\ && \qquad \qquad \qquad \qquad \qquad 
-2 |X| \dot z^\alpha(T) e^\beta{}_i(T) \dot z^\gamma(T) e^\delta{}_j(T) 
\Bigr)
\nonumber \\ && 
-2 V_{\alpha\beta\gamma}(T) e^\alpha{}_i(T) e^\beta{}_j(T) X^\gamma(T)
+O((Gm)^2 |X|/L^3) \,,
\\
{}^{(0)}_{(2)}\bar h_{00} &=& {(Gm)^2 \over|X|^2} \,,
\\
{}^{(0)}_{(2)}\bar h_{0i} &=& -{2 G m \over |X|^3} 
S_{\alpha\beta}(T)X^\alpha(T)e^\beta{}_i(T) +O((Gm)^3/L|X|^2)\,, 
\\
{}^{(0)}_{(2)}\bar h_{ij} &=& {(Gm)^2 \over |X|^2} 
\left(-2 \delta_{ij} +{X^i X^j\over |X|^2}\right) \,,
\\
{}^{(1)}_{(2)}\bar h_{ab}&=& 0 \,.
\end{eqnarray}


\section{Equation of Motion} 
We apply Eqs.~(2.2a), (2.2b), (2.3a) and (2.3b) of \cite{Thorne1} 
to derive the equation of motion, which are
\begin{eqnarray}
P^a(T,r) &=& {1\over 16\pi G}\int_{|X|=r}d^2 S_j H^{ab0j}{}_{,b} 
\label{eq:momentum} \\ 
J^{ij}(T,r) &=& {1\over 16\pi G}\int_{|X|=r}d^2 S_k 
\left(X^i H^{ja0k}{}_{,a}-X^j H^{ia0k}{}_{,a}+H^{ik0j}-H^{jk0i}\right)
\label{eq:spin} \\ 
{d\over dT}P^a(T,r) &=& -{1\over G}\int_S d^2 S_j t^{aj}(X)
\label{eq:force} \\ 
{d\over dT}J^{ij}(T,r) &=& -{1\over G}\int_S d^2 S_k 
\left(X^i t^{jk}(X) -X^j t^{ik}(X) \right)
\label{eq:torque}
\end{eqnarray}
As in \cite{Thorne1}, we adopt the choice of Landau-Lifshitz 
for $H^{abcd}$ and $t^{ab}$. 
\begin{eqnarray}
H^{abcd} &=& H^{abcd}_{L-L}\,,
\\
t^{ab} &=& (-g)t^{ab}_{L-L}\,.
\end{eqnarray}
The explicit expressions are given in standard texts, 
such as, Sec.20 of Misner-Thorne-Wheeler\cite{MTW} or 
Sec.96 of Landau-Lifshitz\cite{LL}.
The integrals are taken over a 2-surface of constant $|X|$ 
in the overlapping region for definiteness, 
but we focus on the $r$-independent terms in the integrals 
since it is enough only to consider them 
in order to derive the leading order correction to the orbit $z(T)$. 

Before considering (\ref{eq:momentum}) and (\ref{eq:force}), 
we first show 
that the spin tensor $S_{\alpha\beta}(T)$ is geodetic transported 
in the background geometry $g_{\alpha\beta}(z(T))$ 
in the test particle limit $m \rightarrow 0$. 
Eq. (\ref{eq:spin}) has a dimension of $(mass)\times(length)$ 
and we consider the terms of order $(Gm)^2$. 
Power counting of $X$ shows that there will be contributions 
linear in the $({}^0_2)$-components of the metric and those from
bilinear combinations of the $({}^0_1)$- and $({}^0_1)$-components
 of the metric. 
We obtain 
\begin{eqnarray}
J^{ij}(T,r) = m S_{\alpha\beta}(T)e^{\alpha i}e^{\beta i} +O(G^2m^3/L)
+(r\mbox{-dependent terms}). 
\label{eq:spin1}
\end{eqnarray}
Eq.~(\ref{eq:torque}) has a dimension of $(mass)^1$ and 
we consider the terms of order $Gm^2/L$ in the same way. 
Power counting of $X$ shows that 
there will be contributions from bilinear combinations of
the $({}^0_1)$- and $({}^0_1)$-components of the metric, 
and we soon find that (\ref{eq:torque}) vanishes. 
\begin{eqnarray}
{d \over dT}J^{ij}(T,r) = O(G^2m^3/L^2) 
+(r\mbox{-dependent terms}) 
\label{eq:torque1}
\end{eqnarray}
Since the spatial triad are geodetic transported 
in the background geometry to the leading order, 
(\ref{eq:spin1}) and (\ref{eq:torque1}) result in 
\begin{eqnarray}
{D S^{\alpha\beta} \over dT}(T) = 0. \label{eq:parallel1}
\end{eqnarray}

We next consider (\ref{eq:momentum}) and (\ref{eq:force}). 
Eq.~(\ref{eq:momentum}) has a dimension of $(mass)^1$ and 
we consider the terms of order $m$ and $Gm^2/L$. 
We find that there will be linear contributions 
from $({}^0_1)$-, $({}^0_2)$-components of the metric, 
and bilinear contributions 
from pairs of $({}^0_1)-$ and $({}^0_1)$-components of the metric. 
We obtain 
\begin{eqnarray}
P^0(T,r) &=& m + O(G^2 m^3/L^2) +(r\mbox{-dependent terms}), \\ 
P^i(T,r) &=& O(G^2 m^3/L^2) +(r\mbox{-dependent terms}). 
\end{eqnarray}
Eq.~(\ref{eq:force}) has a dimension of $(mass)/(length)$ and 
we consider the terms of order $m/L$ and $Gm^2/L^2$. 
There will be bilinear contributions from 
pairs of the $({}^0_2)-$ and $({}^2_0)$-components 
and pairs of the $({}^0_1)-$ and $({}^2_1)$-components of the metric. 
The former pairs give the leading correction to the motion 
due to the spin of the particle, 
and the latter pairs give the radiation reaction of motion. 
A straightforward computation results in 
\begin{eqnarray}
{d\over dT}P^0(T,r) &=& O(G^2 m^3/L^3) +(r\mbox{-dependent terms}), 
\\ 
{d\over dT}P^i(T,r) &=& 
{m\over 2}R_{\alpha\beta\gamma\delta}(z(T))
e^\alpha{}_i(T)\dot z^\beta(T)S^{\gamma\delta}(T)
\nonumber \\ && 
-{m \over 2}V_{\alpha\beta\gamma}(T)e^\gamma{}_i(T)
\left(2\dot z^\alpha(T)\dot z^\beta(T)+g^{\alpha\beta}(z(T))\right)
\nonumber \\ && 
+m\dot z^\alpha(T) {D\over dT}e_{\alpha i}(T) 
+O(G^2 m^3/L^3) +(r\mbox{-dependent terms}). 
\end{eqnarray}
Taking the $T$-derivative of (\ref{eq:Ltt}) and (\ref{eq:Lti}), 
we finally get the equation of motion, 
\begin{eqnarray}
{D \over dT}\dot z^\alpha(T) &=& \bar V_{\beta\gamma\delta}(T) 
\left(2\dot z^\beta(T)g^{\alpha\gamma}(z(T))\dot z^\delta(T) 
-\dot z^\beta(T)\dot z^\gamma(T)g^{\alpha\delta}(z(T))\right) 
\nonumber \\ && 
+{1\over 2}R^\alpha{}_{\beta\gamma\delta}(z(T))
\dot z^\beta(T)S^{\gamma\delta}(T)
+O(G^2 m^2/L^3), \label{eq:result0} 
\end{eqnarray}
where we have defined $\bar V_{\alpha\beta\gamma}=V_{\alpha\beta\gamma}
-1/2 g_{\alpha\beta} V^\delta{}_{\delta\gamma}$. 
Introducing the proper time $\tau$ of the orbit, 
\begin{eqnarray}
{d\tau \over dT}&=& 
1-\bar V_{\alpha\beta}(T)\dot z^\alpha(T)\dot z^\beta(T), 
\end{eqnarray}
 we finally get 
\begin{eqnarray}
{D \over d\tau}{d z^\alpha \over d\tau}(\tau) &=& 
\bar V_{\beta\gamma\delta}(\tau) 
\left({d z^\alpha \over d\tau}{d z^\beta \over d\tau}
{d z^\gamma \over d\tau}{d z^\delta \over d\tau}
+2{d z^\beta\over d\tau}g^{\alpha\gamma}(z(\tau)){d z^\delta\over d\tau} 
-{d z^\beta\over d\tau}{d z^\gamma\over d\tau}g^{\alpha\delta}(z(\tau))
\right) 
\nonumber \\ && 
+{1\over 2}R^\alpha{}_{\beta\gamma\delta}(z(\tau))
{d z^\beta\over d\tau}S^{\gamma\delta}(\tau)
+O(G^2 m^2/L^3). \label{eq:result}
\end{eqnarray}


\section{Conclusion} 
In this paper, we have derived the equation of motion of 
a spinning particle on a given background spacetime. 
We have constructed the metric 
using the technique of matched asymptotic expansion
and integrated the conservation laws introduced in \cite{Thorne1}
to extract the information of the equations of motion. 
One possible drawback of the present derivation of the equation of
motion could be that the consistency of the matched asymptotic 
expansion has not been explicitly demonstrated. To do so, we
need to extract the translational gauge modes from the metric
perturbation in the internal scheme and examine if the condition
of their disappearance leads to the equation of motion,
as done in the previous derivation of the equation of motion for
a non-spinning particle in \cite{Mino1}.
Unfortunately, as mentioned in Introduction, since we have no theory
of the metric perturbation in the Kerr background, it is at present
impossible for us to perform this procedure. However, since we find 
no logical flaw in the method of using the conservation laws, 
we believe the present derivation of the equation of motion is 
perfectly legitimate.

The conclusion of our previous paper\cite{Mino1} is that 
the non-spinning particle moves 
along a geodesic of the regularized perturbed space, 
\begin{eqnarray}
\tilde g_{\mu\nu}(x)=g_{\mu\nu}(x)+h_{(v)\mu\nu}(x), 
\end{eqnarray}
where $h_{(v)\mu\nu}(x)$ is the tail part of the metric perturbation 
(\ref{eq:metper}), 
\begin{eqnarray}
h_{(v)\mu\nu}(x) &=& 
\psi_{(v)\mu\nu}(x)-{1\over 2}g_{\mu\nu}(x)g^{\xi\rho}(x)\psi_{(v)\xi\rho}
(x),
\\
\psi_{(v)\mu\nu}(x)&=&
-2Gm\int^{T_{Ret}(x)}_{-\infty}dT 
\left(v^{\mu\nu}{}_{\alpha\beta}(x,z(T))
\dot z^\alpha(T)\dot z^\beta(T)
+v^{\mu\nu}{}_{\alpha\beta;\gamma}(x,z(T))
S^{\gamma\alpha}(T)\dot z^\beta(T)\right). 
\end{eqnarray} 
We have found
that the spin of the particle gives rise to an effective force 
in addition to this, 
which is the same as derived in \cite{Papa,Dixon,Thorne1}. 
Hence in terms of the regularized metric $\tilde g_{\mu\nu}$
and renormalized proper time $\tilde\tau$, the equation of motion
coincides with the one derived in \cite{Papa,Dixon,Thorne1}:
\begin{eqnarray}
{\tilde D \over d\tilde\tau}\dot z(\tilde\tau) = 
{1\over 2}R^\alpha{}_{\beta\gamma\delta}(z(\tilde\tau))
\dot z^\beta(\tilde\tau)S^{\gamma\delta}(\tilde\tau)\,.
\end{eqnarray}
As was commented in \cite{Mino1}, 
it is still under investigation to find a method
to explicitly compute $h_{(v)\mu\nu}(x)$ even for well-known background
 geometries, such as a Kerr geometry. 


\acknowledgments
We thank all the members of the theoretical gravitational-wave physics
group in Japan for fruitful conversations. 
Special thanks are due to M. Shibata for invaluable discussions.
YM thanks Prof. H. Sato and Prof. S. Ikeuchi for their 
continuous encouragements. This work was supported in part 
by Monbusho Grant-in-Aid for Scientific Research No.5427
and No.08NP0801.

\end{document}